# The *3d* Transition Metal Single Atom Catalysis for Prominent Adsorption Performance Towards Cr(VI) : A First Principle Study

Jianglu Ping[1,] *, Haojun Jia[2]

[1]Northfield Mount Hermon School, MA, 01354, United States;
[2]Department of Chemistry, Massachusetts Institute of Technology, MA, 02139, United States.
**\*Corresponding Author:** email: jianglu4517@gmail.com

## ABSTRACT

The reduction reaction of turning Cr(VI) into Cr(III) by appropriate oxidants together with nanoscale catalysts in contaminated soil and water received significant amount of attention for its environmental benefits. Single-atom catalysts (SACs), as an emergent new type of catalysts, show great potential for simultaneous adsorption for the Cr(VI) removal. In this work, we investigated the efficiency of 3*d* transitional metal (from Sc to Zn) SACs in adsorption for the Cr(VI) removal via a theoretical approach. DFT calculations were used to measure the structural and energetic of Cr(VI) ($H_2CrO_4$ and $Cr_2O_7^{2-}$) adsorption behavior on the SACs systems. The Ti-N-C has a good performance for the $H_2CrO_4$ and $Cr_2O_7^{2-}$ removal and has important environmental protection significance for water purification. In addition, our results show that the binding energy of the SACs systems for $H_2CrO_4$ adsorption is found to be much less than that for $Cr_2O_7^{2-}$ adsorption, indicating that reduction of Cr(VI) in $Cr_2O_7^{2-}$ is energetically favorable than that of Cr(VI) in $H_2CrO_4$.

# INTRODUCTION

Unsuitable treatments approaches to landfills have caused chromium contamination to soil and ground water [1]. In natural environment, Chromium is mainly found with two oxidation states: Cr(VI) and Cr(III). On the contrary of Cr(III), Cr(VI) is highly soluble, mobile and toxic to humans, animals and plants [2]. Chromium is non-essential element for plants and there is no uptake mechanism, hence, Cr(VI) is the most persistent element in nature and it cannot be absorbed naturally. Cumulating chromium in plant tissues leads to toxic effects such as leaf chlorosis, poor yield, stunted growth, and less development of root systems. According to Wilson et al., [3], the toxicity of chromium reduces enzyme activities relating to starch and nitrogen metabolism as it hinders relevant enzymes or produces reactive oxygen species that are fatal to plants. On the contrary of Cr(VI), Cr(III) is found to have low toxicity, as a crucial trace metal in human nutrition, and is generally found as $Cr(OH)_3$ or Cr(III) complexes with organic ligands [4]. The study of Costa et al. [5] indicates that while Cr(VI) suppresses viability and chlorophyll autofluorescence, Cr( III) processed higher stimulatory impact on reactive oxygen species formation and lipid peroxidation. The concentration of chromium in drinking water has been regulated legislatively to be less than 0.05 mg $L^{-1}$ in many countries such as Germany, Sweden Argentine, and China [6]. As a result, eliminating Cr(VI) is essential to environmental security and the reduction reaction of turning Cr(VI) into Cr(III) in contaminated soil and water received significant amount of attention for its environmental benefits.

Cr(VI) reduction by appropriate oxidants together with nanoscale catalysts is considered as catalytic reduction [7]. The natural Cr(VI) reduction reaction proceeds with a slow reaction rate, and the addition of nanoparticles, reductants, or other catalysts could increase its reaction rate. Platinum groups metals that do not carry toxic effects to water are generally put in. Since the platinum group metals are scarce to find, utilizing these metals to catalyze the reduction reaction are not practical and economical. Instead of platinum metals, transition metals, such as Mn(II) and Fe(III), could be widely used in homogeneous and heterogeneous catalysts to increase the reaction rate of Cr(VI) reduction [8, 9]. Such reactions are consisted of an initially slow inducing reaction stage and a final accelerating reaction stage. Recently, photocatalytic impacts of Cr(VI) reduction by organic acids received considerable attention from researchers [4]. Previous work [9] show that Fe(III) with oxalic acid and citric acid can catalyze the photochemical reduction reaction of Cr(VI). Furthermore, Cr(VI) photocatalytic reduction with $TiO_2$ and organic compounds received a significant amount of attention as well [9]. In this catalytic reaction, rapid removal of Cr(VI) by organic compounds with different numbers of carboxylic groups could be observed.

Single-atom catalysts (SACs), as an emergent new type of catalysts, contain isolated single metal atoms dispersed on supports, which are frequently graphitic [10]. The structure and properties of SACs significantly differs from those of nanoparticles and metal clusters, allowing for greater activity and durability for various catalytic reactions. SACs capture the superiority of both homogeneous and heterogeneous catalysts by combining active site tunability with scalability [11]. The unique electronic structure and unsaturated coordination environments of the active centers in SACs have been proven to improve catalytic activity in a variety of reactions [12]. The metal atoms in SACs always display a distinctive HOMO-LOMO gap and a discrete energy level distribution, giving them a unique energy level structure. It is worth noting that due to the atomic dispersion configuration in SACs, it is likely to attain a theoretical efficiency of 100% in catalytic reactions, which greatly enhances the usage of the metal atoms. The development of SACs can enable the reasonable use of metal resources and facilitate atomic economy through recycling and maximum atom-utilization efficiency. SACs that have geometric structures with all active sites exposed and accessible are found to have catalytically efficient structures. The practical utilization efficiency of active metal sites in SACs may be less than 100% even though the metal atom utilization may be 100% because some of the active metal sites may be embedded. Numerous methods have been developed to improve the structure of SACs in order to build more attainable interfaces and enable effective usage of the active metal sites. The maximum atom-utilization efficiency of SACs makes them exhibit excellent performance at a low consumption, which is beneficial for the decrease of catalyst cost. It is well accepted that increasing the number of active sites and enhancing the intrinsic activity of each active site are two aspects for boosting the activity of catalysts [13-15]. SACs with fully exposed active atoms can effectively increase the number of active sites. The low-coordination environment and charge-transfer effect of SACs via enhanced metal-support interaction has been proven to enhance the intrinsic activity of active sites [16]. Based on the superiority of SACs, employing single-atom materials for Cr(VI) reduction is a promising approach to reduce catalyst cost and enhance catalytic performance for Cr(VI) reduction.

It is common knowledge that each element has unique chemical characteristics. When a single metal atom is assumed to be the active site, various single metal atoms with various atomic configurations may exhibit different behaviors. Sometimes the active component might be thought of as the support for a single metal atom. Through the potent connection between the metal and the support, doping individual metal atoms would alter the activity of the matrix. As a result, this interaction may differ for various metal SACs on the same support, leading to various activities. Doping various metal atoms allows for the modulation of both the structural and electrical properties. One of the critical elements to achieving high activity for both HER and OER may be the proper element selection for the doping metal. Even so, there are too many different SAC variants to look into individually. Correlations between the atomic structures of the doping metal and the support must be examined in order to understand how they might impact the SAC's activity.

Graphene is a novel material which was first prepared in 2004 [17]. The carbon atoms in graphene are connected by *sp²* hybrids and closely packed into a single-layer two-dimensional honeycomb lattice structure [17]. Graphene has become a hot topic in material [18-19], energy [20-22] and other disciplines. Given the unique properties, such as optical [23], magnetic [24], luminescent [25], electronic and thermal transport properties [26], researchers can obtain advanced functional materials when graphene is used or in combination with other materials for sensors [27], bioimaging [28], electronics [29] and photovoltaics [30]. In addition, graphene is an ideal two-dimensional (2D) catalyst support because of its large surface area which allows it to have novel physical and chemical properties [31]. Metal species in SACs are all downsized into nanoparticles and are located dispersedly on graphene, which increases the catalytic efficiency for the more exposed unsaturated reactive sites [32]. Typically, the metal atom in the SACs is anchored to the N sites graphene support [33]. N is an ideal heteroatom dopant because its atomic radii is similar to that of for C atom and its notable electronegativity difference comparing with C atom. For applications requiring SACs, nitrogen-doped carbon stands out among all the carbon materials [34]. Both experimental and density functional theory (DFT) calculation have proved that N is a more ideal doping type than graphitic because of its weaker binding energy [34]. The electric neutrality of the carbon matrix is effectively disrupted by nitrogen doping. As a result, the more electronegative nitrogen atoms align with the single metal atoms to form an MNx structure, which has been confirmed to be the active center for various electrochemical reactions [13-16]. Therefore, numerous carbon-based SACs containing a single MNx site have been extensively used. Since graphene-based SACs is highly active and selective, graphene-based SACs can be applied to various purposes and have become a research hotspot in various fields such as energy and environmental protection [35].

Simultaneous adsorption is ideal reduction method for Cr(VI) removal because Cr(VI) has strong oxidation and high toxicity. Treatment technologies such as adsorption–reduction technology with appropriate adsorbent is essential for the removal of Cr(VI). SACs, having high performing redox and adsorption properties, are advantageous for the effective removal of heavy metals [36]. Due to a lack of control over SACs site configuration, relationships between the structure of SACs active site and Cr(VI) adsorption site are challenging to deconvolute via experiment. Moreover, because the sub-nm scale can challenge even the highest-resolution spectroscopic probes that are sensitive to local variations in local chemistry connectivity, the reactivity and selectivity of SACs are poorly understood from an experimental perspective. Computational modeling enables us to study Cr(VI) adsorption configurations of SACs with atomic precision and can elucidate the removal mechanism of SACs on reduction Cr(VI) in nanoscale from both thermodynamics and kinetics.

Motivated by the importance of the reduction reaction of turning Cr(VI) into Cr(III) in contaminated soil and water, in this work, we theoretically investigate the Cr(VI) ($H_2CrO_4$ and $Cr_2O_7^{2-}$) adsorption behavior on SACs. Herein, we calculate the *3d* transition metal (from Sc to Zn) SACs via DFT to predict the stability and affinity of SACs with the two formats of Cr(VI) and electronic properties. Our simulation can screen all the theoretical models efficiently and give guidance to experiment researchers.

# COMPUTATIONAL METHOD

In this work, the 3d transition metal (from Sc to Zn) SACs and with corresponding $H_2CrO_4$ and $Cr_2O_7^{2-}$ adsorption were calculated. Gas phase geometry optimization calculations are computed by DFT implemented with quantum chemistry electronic structure code ORCA package [37]. The functional B3LYP and basis set lacvps_ecp are used in all the calculations. The default threshold SCF convergence is set as the value of $3 \times 10^{-5}$.

$$\Delta E_{binding\ energy} = E_{SACs-molecule} - E_{SACs} - E_{molecule} \quad (1)$$

To discuss binding affinity of the SACs towards $H_2CrO_4$ and $Cr_2O_7^{2-}$, $E_{SACs\text{-}molecule}$, $E_{SACs}$, and $E_{molecule}$ are introduced, where $E_{SACs\text{-}molecule}$ is the total energy of SACs binding with $H_2CrO_4$ or $Cr_2O_7^{2-}$, $E_{SACs}$ is the total energy of SACs and $E_{molecule}$ is the energy of $H_2CrO_4$ or $Cr_2O_7^{2-}$ in gas phase condition with the same functional and basis set.

# RESULTS AND DISCUSSION

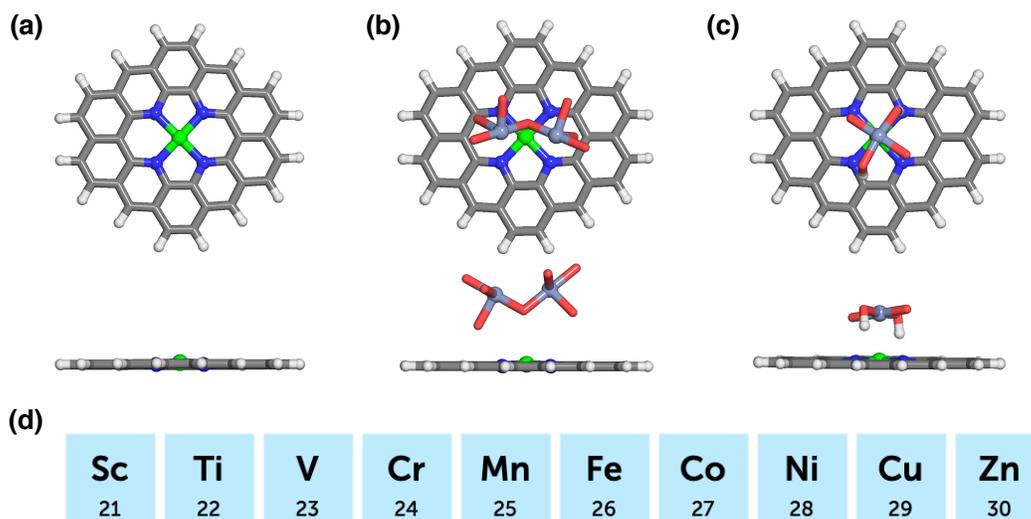

**Figure 1,** Initial atomic structures of *3d* transition metal-embedded (from Sc to Zn) N, O, P and S-doped graphene SACs is demonstrated in (a). The $Cr_2O_7^{2-}$ and $H_2CrO_4$ molecules adsorption on SACs are shown in (b) and (c), respectively. The metal system investigated in this work are shown in (d). The atomic structures are shown in the ball-and-stick representation colored as follows: metals (from Sc to Zn) in green, dopants (N, O, P and S) in blue, Cr in blue-gray, O in red, C in gray, and H in white.

To study the Cr(VI) adsorption effects on the *3d* transition metal-embedded (from Sc to Zn) SACs and its thermodynamical stability, DFT calculations were employed to reveal the interaction mechanism between SACs and Cr(VI). The two most common existence forms of Cr(VI) ($Cr_2O_7^{2-}$ and $H_2CrO_4$ molecules) were considered in this work. In addition, different direct metal-coordinating environment (i.e., N, O, P, S) of SACs is investigated to understand the influence of dopants on the structural and energetic properties (Figure 1). In our work, we comprehensively studied *3d* transition metal (from Sc to Zn) finite graphitic SAC models with identical metal-coordinating dopants (i.e., N, O, P, S) in 6-membered rings within the graphene model. First, the 57-atom graphitic SACs flake model is obtained from a data set of structures that were optimized with ωPBEh/ LACVP* [11]. We used this initial model to construct our following $MX_4C_{10}$ model (where M = Sc, Ti, V, Cr, Mn, Fe, Co, Ni, Cu, Zn; X= N, O, P, S) with the chemical formula C$_{36}$ X$_4$H$_{16}$M that corresponds to all metal-coordinating atoms in 6-membered rings. N, O, P or S-doped graphene SACs are used to investigate how the coordinating dopants effect on the interactions of H$_2$CrO$_4$ and Cr$_2$O$_7^{2-}$ with transition metal SACs.

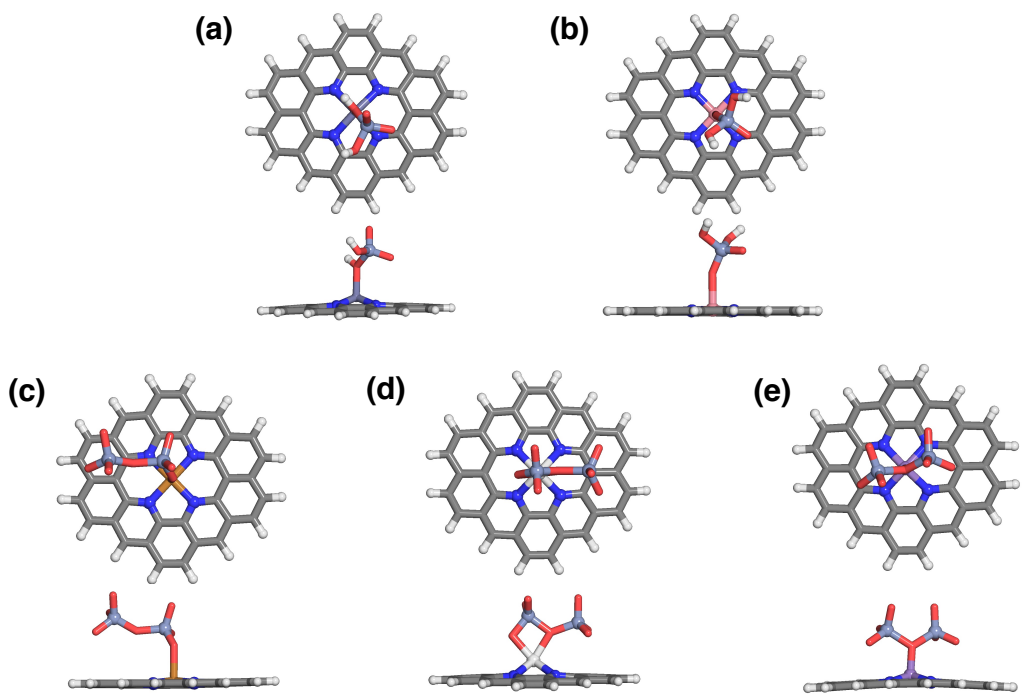

**Figure 2,** The configurations of H$_2$CrO$_4$ and Cr$_2$O$_7^{2-}$ adsorption on the N, O, P or S-doped graphene transition metal SACs. For H$_2$CrO$_4$ adsorption on transition metal SACs, two configurations as demonstrated in (a) and (b). For Cr$_2$O$_7^{2-}$ adsorption on transition metal SACs, three possibilities as show in (c)-(e). The atomic structures are shown in the ball-and-stick representation colored as follows: Zn in galaxy-gray, Co in pink, Cu in orange, Sc in sliver-white, Mn in purple, N in blue, Cr in blue-gray, O in red, C in gray, and H in white.

For H$_2$CrO$_4$ adsorption on transition metal SACs, there are two types of adsorption models. As shown in Figure2 (a), one type of the model has their central metal atom binding with OH and Mn SACs with H$_2$CrO$_4$ adsorption is in this configuration. As shown in Figure 2 (b), the transitional metal atoms in another type of H$_2$CrO$_4$ adsorption model is directly binding with O, which includes Co, Cr, Cu, Fe, Ni, Sc, Ti, V, Zn SACs systems. For Cr$_2$O$_7^{2-}$ adsorption on the N, O, P, or S doped graphene transition metal SACs, there are three different kinds of configurations. SACs Systems with transition metal atoms binding to one oxygen termination of Cr$_2$O$_7^{2-}$, such as Cu-SACs (Figure2 c). In contrast, Cr, Sc, Ti, V SACs binding to two oxygen termination (Figure2 d), and Co, Fe, Mn, Ni, Zn SACs are connected to the oxygen in the middle (Figure2 e). For convenience, SACs system discussed above are all doped with N. Doping with different element would not affect the SAC's adsorption model, and the adsorption models are in fact mainly determined by the identity of the central metal atom. For SACs systems that are doped with P and S, O is likely to insert with P and S dopants, which causes metal dopant bond lengths for these systems to be larger than those for N, O doped systems.

Average metal-dopant bond lengths (in Å) in the SAC systems differ based on the dopants. In N-doped SACs, the maximum average metal-dopant bond length is 2.11Å, belonging to Sc-N-C, and the minimum average metal-dopant bond length is 1.90 Å, belonging to Ni-N-C. The average metal-dopant bond length for N-doped SACs follows this trend: Sc> Ti, V, Zn> Cr> Mn, Cu> Fe> Co> Ni. After $H_2CrO_4$ adsorption, most SACs models demonstrated change in average metal-dopants bond lengths. In N-doped SACs, average metal-dopants bond lengths in Zn-N-C exhibited the greatest change in bond lengths as it decreased by 0.45Å. In O-doped SACs, Co-O-C has the greatest increase in average bond length after $H_2CrO_4$ adsorption by 0.25 Å. In P-doped SACs, Cr-P-C is found to have the largest decrease in average metal-dopant bond length by 0.12 Å. For S- doped SACs systems, average metal-dopants bond lengths of Co-S-C increased by 0.31 Å. Apart from SACs systems that showed change in average metal-dopant bond lengths, there are three systems that showed no change in average bond length: Co-N-C, Ni-O-C, Cu-P-C. We concluded that these systems are not sensitive to $H_2CrO_4$ adsorption. After $Cr_2O_7^{2-}$ adsorption, four SACs systems are found to be insensitive to the adsorption as their metal-dopant bond lengths changed by 0 compared to the ones after $H_2CrO_4$ adsorption. These systems are: Ti-N-C, Mn-N-C, Ni-N-C, Cu-N-C. Aside from these, average metal-dopant bond lengths of Zn-N-C increase for 0.59Å, which is the largest increase in average metal-dopant bond lengths in N-doped SACs. In

O-doped SACs, the system embedded with Ni is found to have the largest decrease in average metal-dopant bond length for 1.45 Å. In P-doped SACs, Cr-P-C showed the largest increase in average metal-dopant bond length for 0.19 Å. For S- doped SACs systems, Cu-S-C is found to have the largest increase in average metal-dopant bond length.

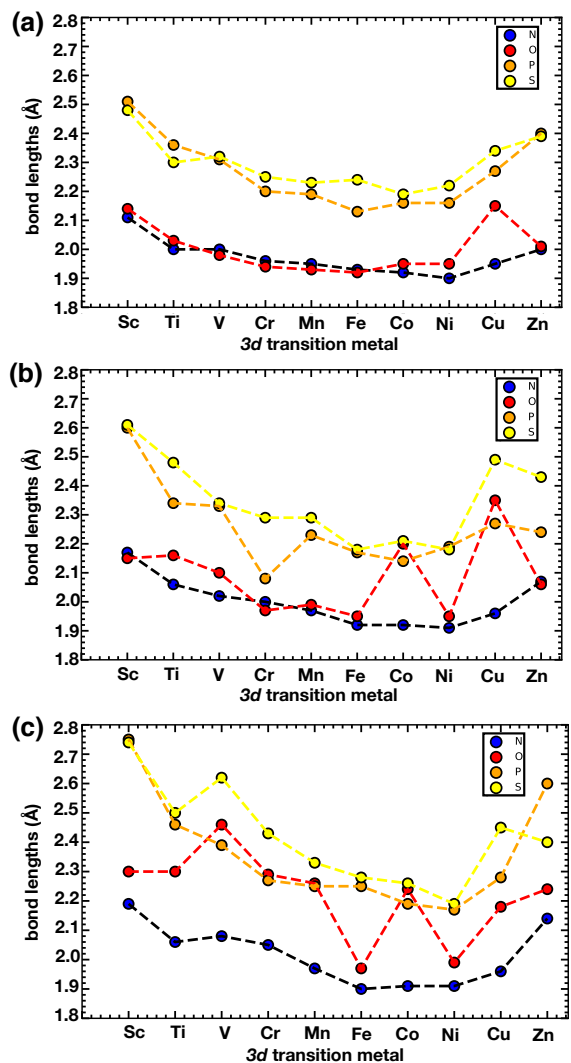

**Figure 3.** Calculated average metal-dopant bond lengths for 3d transition metals (Sc, Ti, V, Cr, Mn, Fe, Co, Ni, Cu and Zn) SACs, SACs binding with and $H_2CrO_4$ and $Cr_2O_7^{2-}$. The N, O, P and S doped SACs are shown in blue, red, orange and yellow, respectively.

In order to investigate the ligan filed dependence on different metals and dopants, we firstly evaluate the average metal-dopant bond lengths among different 3d transition metal SACs with various local chemistry environments. The average metal-dopants bond length of *3d* transition metal-embedded (from Sc to Zn) N, O, P, S-doped SACs is shown in Figure 3(a), which indicates that the average metal-dopants bond length when graphene is doped with P or S is larger than that is doped with N or O. Average metal-dopants bond length for SACs models doped with N, O, P, or S ranges from 1.90-2.11 Å, 1.92- 2.14 Å, 2.13- 2.51 Å, and 1.9- 2.48 Å. These qualities are metal dependent and not influence by dopant identities. When the SACs systems were embedded with transition metals such as Cr, Mn, Fe, Co, and Ni, the average metal-dopants bond length of the systems were found to be shorter than those embedded with Sc, Ti, V, Cu, or Zn.

After $H_2CrO_4$ adsorption, the average metal-dopant bond length in the SACs models we studied increased in general. Shown in Figure 3b), the average bond length Co-O-C and Cu-O-C exhibited large increases each by 0.25 Å and 0.20 Å. Given the same metal, the average metal-dopants bond length is the shortest when the graphene is doped with N, except for Sc, Cr, Zn. It is worth noting that the average metal-dopants bond length for some O-doped SACs were greater than P-doped or S- doped SACs when embedded with the same metal. For instance, metal-dopants bond lengths for Co-O-C and Cu-O-C systems became greater than that for Co-P-S and Cu-P-S. After $Cr_2O_7^{2-}$ adsorption (Figure 3 c)), the metal-dopants bond lengths for most metals embedded on N, O, S, P doped graphene increased. After $Cr_2O_7^{2-}$ adsorption, the average metal-dopant bond length in the SACs models we studied increased in general, compared to the average metal-dopants bond length each system exhibited after $H_2CrO_4$ adsorption, except that there are a few that decreased. For instance, Ni-O-C system is found to have a decrease of 1.45 Å in its average metal-dopants bond length. Apart from systems like Ni-O-C, Cu-S-C system is found to have an obvious increase of 1.21 Å in its average metal-dopants bond length after $Cr_2O_7^{2-}$ adsorption. Overall, the average metal-dopants bond length in the SACs systems increased after adsorptions with $H_2CrO_4$ and $Cr_2O_7^{2-}$. Among all transition metal embedded N, O, P, S doped graphene SACs, N-doped

SACs are found to have the least metal-dopants bond length, which indicates N-doped SACs have the strongest metal-ligand field among N, O, P and S doped systems.

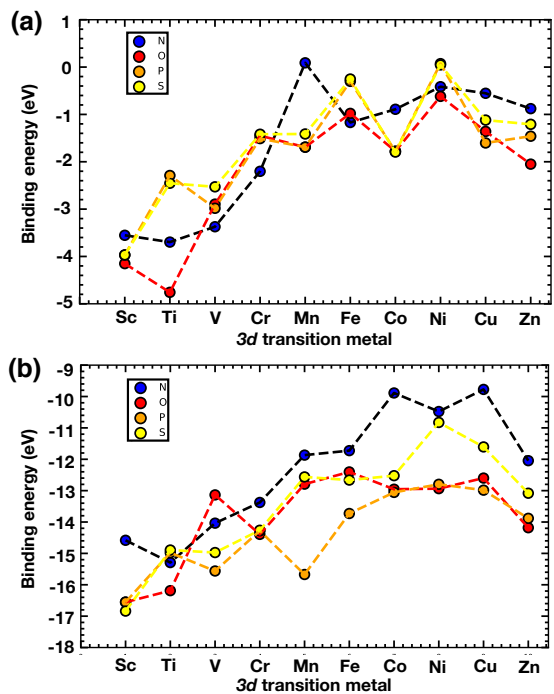

**Figure 4.** $H_2CrO_4$ and $Cr_2O_7^{2-}$ binding energy with 3d transition metals (Sc, Ti, V, Cr, Mn, Fe, Co, Ni, Cu and Zn) SACs. The N, O, P and S doped SACs are shown in blue, red, orange and yellow, respectively.

In order to further clarify the interaction mechanism between the SACs and $H_2CrO_4$ and $Cr_2O_7^{2-}$, we make the energetic comparison of $H_2CrO_4$ and $Cr_2O_7^{2-}$ adsorption among 3 d transition metals with different coordinating environment. As shown in Figure 4a), the binding energy of $H_2CrO_4$ with 3d transition metal SACs which are doped with N, O, P, or S ranges from -4.8eV to 0.3eV. The Sc, Ti, and V embedded SACs are found to have lower binding energies with $H_2CrO_4$ in general in different coordinating environment. The binding energy of $Cr_2O_7^{2-}$ with 3d transition metal SACs which are doped with N, O, P, or S varies from -16.9eV to -9.8eV. The figure indicates that Sc and Ti embedded SACs tend to have lower binding energy with $Cr_2O_7^{2-}$. Figure 4 b) also shows that when 3d transition metals SACs are doped with N, their binding energy with $Cr_2O_7^{2-}$ is higher than those doped with O, P, or S. In contrast, SACs models doped with P generally have lower binding energy with $Cr_2O_7^{2-}$. The adsorption energetics is also highly dependent on metal identity. SACs models embedded with Sc, Ti, V, and Cr all have relatively lower binding energies regardless of dopants and adsorption stage. In contrast, SACs models embedded with transition metals such as Mn, Fe, Cu, Ni, Co, and Zn are found to have higher binding energies irrespective of dopant identity and adsorption stage. Furthermore, the $Cr_2O_7^{2-}$ adsorption on SACs is significantly favorable comparing with $H_2CrO_4$ adsorption behaviors. $H_2CrO_4$ and $Cr_2O_7^{2-}$ binding energy with 3d transition metals SACs are much energetically favorable than $H_2CrO_4$ and $Cr_2O_7^{2-}$ interaction with nitrogen-doped $KFeS_2/C$ composite [38], which demonstrates the promising catalytic performance of SACs.

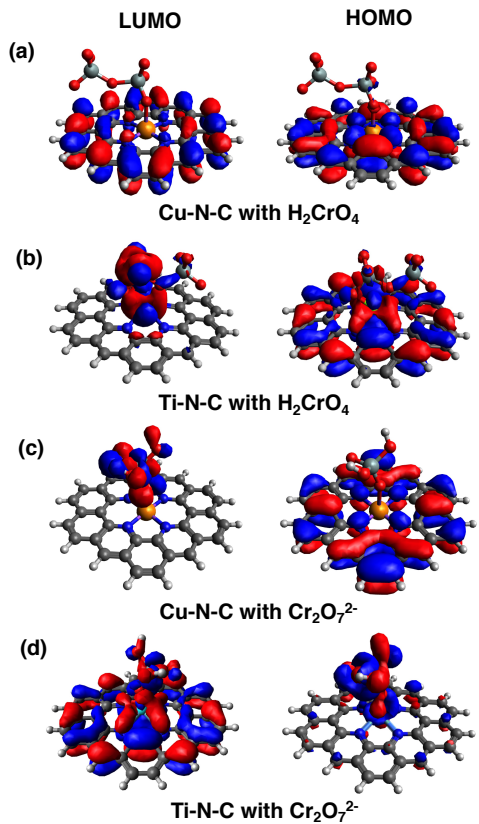

(a) Cu-N-C with $H_2CrO_4$
(b) Ti-N-C with $H_2CrO_4$
(c) Cu-N-C with $Cr_2O_7^{2-}$
(d) Ti-N-C with $Cr_2O_7^{2-}$

**Figure 5.** HOMO and LUMO electron distribution of Cu-N-C and Ti-N-C systems after $H_2CrO_4$ adsorption and $Cr_2O_7^{2-}$ adsorption. The positive and negative density of electron are shown in red and blue, respectively. The isosurface value of 0.02 was used for all the systems.

Figure 5 demonstrates the spatial distribution of HOMO and LUMO energy levels of Cu-N-C and Ti-N-C systems after $H_2CrO_4$ adsorption and $Cr_2O_7^{2-}$ adsorption, respectively. Cu-N-C is chosen here because it performed relatively high binding energy in both $H_2CrO_4$ adsorption and $Cr_2O_7^{2-}$ adsorption, while Ti-N-C system is an example of systems with relatively low binding energy in the two adsorption processes. There are both vertical and horizontal symmetries in the HOMO of Cu-N-C after $H_2CrO_4$ adsorption, whereas its LUMO only has horizontal symmetry. In addition, comparing with spatial distribution of $H_2CrO_4$ adsorption, positive and negative electron distribution in the HOMO of Cu-N-C after $Cr_2O_7^{2-}$ adsorption showed an obvious decrease, while its LUMO displayed an increase in electron distribution. For Ti-N-C after both $H_2CrO_4$ adsorption and $Cr_2O_7^{2}$, the LUMOs are principally located on the Ti atoms.

# CONCLUSION

In this work, we investigated the efficiency of 3$d$ transitional metal (from Sc to Zn) SACs in Cr(VI) reduction reaction via a theoretical approach. We used DFT calculation to study the Cr(VI) ($H_2CrO_4$ and $Cr_2O_7^{2-}$) adsorption behavior on the SACs systems. The average metal-dopants bond length in the SACs systems increased after adsorptions with $H_2CrO_4$ and $Cr_2O_7^{2-}$. The N-doped SACs are found to have the least metal-dopants bond length, which indicates N-doped SACs have the strongest metal-ligand field among N, O, P and S doped systems. Our simulation indicated that among all 3$d$ transitional metals, Ti is the best candidate for the central metal atom of SACs for Cr(VI) reduction. On the other hand, the performance of metals such as Cu, Co, Zn are less ideal in Cr(VI) removal. In addition, our simulation also shows that reduction of Cr(VI) in $Cr_2O_7^{2-}$ is energetically favorable than that of Cr(VI) in $H_2CrO_4$.

**Conflicts of Interest:** The authors declare no conflict of interest.

# REFERENCE

[1] M. Shahid, S. Shamshad, M. Rafiq, S. Khalid, I. Bibi, N. K. Niazi, C. Dumat, M. I. Rashid, Chromium speciation, bioavailability, uptake, toxicity and detoxification in soil-plant system: a review. *Chemosphere*, **2017**, 178, 513–533.

[2] T. K. Jan, D. R. Young, Chromium speciation in municipal wastewaters and seawater. *J. Water Pollut. Control Fed.*, **1978**, 2327–2336.

[3] W. Wilson, Q. Zhang, R. E. M. Rickaby, Susceptibility of algae to Cr toxicity reveals contrasting metal management strategies. *Limnol. Oceanogr.*, **2019**, 64, 2271–2282.

[4] J. Sun, J.-D. Mao, H. Gong, Y. Lan, Fe(III) photocatalytic reduction of Cr(VI) by low-molecular-weight organic acids with α-OH. *J. Hazard. Mater.*, **2009**, 168, 1569–1574

[5] C. H. Costa, F. Perreault, A. Oukarroum, S. P. Melegari, R. Popovic, W. G. Matias, Effect of chromium oxide (III) nanoparticles on the production of reactive oxygen species and photosystem II activity in the green alga Chlamydomonas reinhardtii. *Sci. Total Environ.*, **2016**, 565, 951–960.

[6] J.J. Testa, M.A. Grela, M.I. Litter, *Environ. Sci. Technol.*, **2004**, 38, 1589.

[7] Z.H. Farooqi, M.W. Akram, R. Begum, W. Wu, A. Irfan, Inorganic nanoparticles for reduction of hexavalent chromium, *J. Hazard. Mater.*, **2020**, 123535.

[8] Kabir-ud-Din, K. Hartani, Z. Khan, One-Step Three-Electron Oxidation of Tartaric and Glyoxylic Acids by Chromium(VI) in the Absence and Presence of Manganese(II). *Transit. Metal Chem.*, **2002**, 27, 617-624.

[9] G. Colon, M. C. Hidalgo, J. A. Navio, Influence of Carboxylic Acid on the Photocatalytic Reduction of Cr(VI) using Commercial Ti02. *Langmuir*, **2001**, 17, 7174-7177

[10] X. F. Yang, A. Wang, B. Qiao, J. Li, J. Liu, T. Zhang, Single-atom catalysts: a new frontier in heterogeneous catalysis. *Acc. Chem. Res.*, **2013**, 46, 1740–1748.

[11] H. J. Jia, A. Nandy, M. Liu, H. J. Kulik, Modeling the Roles of Rigidity and Dopants in Single-Atom Methane-to-Methanol Catalysts. *J. Mater. Chem. A*, **2022**, 10, 6193– 6203

[12] G. Liu, A. W. Robertson, M. M. Li, W. C. H. Kuo, M. T. Darby, M. H. Muhieddine, Y. C. Lin, K. Suenaga, M. Stamatakis, J. H. Warner, S. C. E. Tsang, MoS2 monolayer catalyst doped with isolated Co atoms for the hydrodeoxygenation reaction. *Nat. Chem.*, **2017**, 9, 810–816.

[13] B. Qiao, A. Wang, X. Yang, L. F. Allard, Z. Jiang, Y. Cui, J. Liu, J. Li, T. Zhang, Single-atom catalysis of CO oxidation using Pt1/FeOx. *Nat. Chem.*, **2011**, 3, 634–641.

[14] L. Lin, W. Zhou, R. Gao, S. Yao, X. Zhang, W. Xu, S. Zheng, Z. Jiang, Q. Yu, Y. W. Li, C. Shi, X. D. Wen, D. Ma, Low-temperature hydrogen production from water and methanol using Pt/a-MoC catalysts. *Nature*, **2017**, 544, 80–83.

[15] A. Wang, J. Li, Y. Zhang, Heterogeneous single-atom catalysis. *Nat. Rev. Chem.*, 2018, 2, 65–81.

[16] Y. Chen, S. Ji, Y. Wang, J. Dong, W. Chen, Z. Li, R. Shen, L. Zheng, Z. Zhuang, D. Wang, Y. Li, Isolated single iron atoms anchored on N-doped porous carbon as an efficient electrocatalyst for the oxygen reduction reaction. *Angew. Chem. Int. Ed.*, **2017**, 56, 6937–6941.

[17] A. K. Geim, K. S. Novoselov, The rise of graphene. *Nature Mater.*, **2007**, 6, 183–191.

[18] R. R. Nair, P. Blake, A. N. Grigorenko, K. S. Novoselov, T. J. Booth, T. Stauber, N. M. R. Peres, A. K. Geim, Fine Structure Constant Defines Visual Transparency of Graphene. *Science.*, **2008**, 320, 1308–1308.

[19] T. Liang, Z. Zhang, X. Feng, H. Jia, C. J. Pickard, S. Redfern, D. Duan. Ternary hypervalent silicon hydrides via lithium at high pressure. *Phys. Rev. Mater.*, **2020**, 4, 113607.

[20] H. J. Jia, H. M. Mu, J. P. Li, Y. Z. Zhao, Y. X. Wu, X. C. Wang, Piezoelectric and polarized enhancement by hydrofluorination of penta-graphene. *Phys. Chem. Chem. Phys.*, **2018**, 20, 26288–26296.
[21] J. P. Li, H. J. Jia, D. R. Zhu, X. C. Wang, F. C. Liu, Y. J. Yang, Piezoelectricity and Dipolar Polarization of Group V-IV-III-VI Sheets: A First-Principles Study. *Appl. Surf. Sci.*, **2019**, 463, 918– 922.
[22] Y. Z. Zhao, H. J. Jia, S. N. Zhao, Y. B. Wang, H. Y. Li, Z. L. Zhao, Y. X. Wu, X. C. Wang, Janus structure derivatives SnP–InS, GeP-GaS and SiP–AlS monolayers with in-plane and out-of-plane piezoelectric performance, *Physica E*, **2020**, 117, 113817
[23] J. Vähäkangas, P. Lantto, J. Vaara, Faraday Rotation in Graphene Quantum Dots: Interplay of Size, Perimeter Type, and Functionalization, *J. Phys. Chem. C*, **2014**, 118, 23996–24005.
[24] Y. Wu, Y. Jiao, Y. Zhao, H. Jia, L. Xu, Noise-induced quasiperiod and period switching, *Phys. Rev. E*, **2022**, 105, 014419
[25] S. Zhu, J. Shao, Y. Song, X. Zhao, J. Du, L. Wang, H. Wang, K. Zhang, J. Zhang, B. Yang, Investigating the Surface State of Graphene Quantum Dots, *Nanoscale*, **2015**, 7, 7927–7933.
[26] K. S. Novoselov, A. K. Geim, S. V. Morozov, D. Jiang, M. I. Katsnelson, I. V. Grigorieva, S. V. Dubonos, A. A. Firsov, Two-Dimensional Gas of Massless Dirac Fermions in Graphene, *Nature*, **2005**, 438, 197–200.
[27] S. Chen, Y. Song, Y. Li, Y. Liu, X. Su, Q. Ma, A Facile Photoluminescence Modulated Nanosensor Based on NitrogenDoped Graphene Quantum Dots for Sulfite Detection, *New J. Chem.*, **2015**, 39, 8114–8120.
[28] P. Miao, K. Han, Y. Tang, B. Wang, T. Lin, W. Cheng, Recent Advances in Carbon Nanodots: Synthesis, Properties and Biomedical Applications, *Nanoscale*, **2015**, 7, 1586–1595.
[29] A. D. Guclu, P. Potasz, P. Hawrylak, Excitonic Absorption in Gate-Controlled Graphene Quantum Dots, *Phys. Rev. B: Condens. Matter Mater. Phys.*, **2010**, 82, 155445.
[30] Y. Qin, Y. Cheng, L. Jiang, X. Jin, M. Li, X. Luo, G. Liao, T. Wei, Q. Li, Top-down Strategy toward Versatile Graphene Quantum Dots for Organic/Inorganic Hybrid Solar Cells, *ACS Sustainable Chem. Eng.*, **2015**, 3, 637–644.
[31] B. Bayatsarmadi, Y. Zheng, A. Vasileff, S.Z. Qiao, *Small*, **2017**, 13, 1700191.
[32] T.T. Sun, L.B. Xu, D.S.Wang, Y.D. Li, *Nano Res.*, **2019**, 12, 2067–2080.
[33] L. Zhao, Y. Zhang, L. B. Huang, X. Z. Liu, Q. H. Zhang, C. He, Z. Y. Wu, L. J. Zhang, J. P. Wu, W. Yang, L. Gu, J-S. Hu, L-J Wan, Cascade anchoring strategy for general mass production of high-loading single-atomic metal-nitrogen catalysts. *Nat. Commun.*, **2019**, *10*, 1278.
[34] W. Y. Fujimoto, S. Saito, Formation, stabilities, and electronic properties of nitrogen defects in graphene, *Phys. Rev. B.*, **2011**, 84, 245446.
[35] J. Lin, A. Wang, B. Qiao, X. Liu, X. Yang, X. Wang, Remarkable Performance of Ir$_1$/FeO$_x$ Single-Atom Catalyst in Water Gas Shift Reaction, *J. Am. Chem. Soc.*, **2013**, 135, 15314.
[36] L. Liu, A. Corma, Metal catalysts for heterogeneous catalysis: From single atoms to nanoclusters and nanoparticles. *Chem. Rev.*, **2018**, 118, 4981–5079.
[37] F. Neese, The ORCA program system, *WIREs Comput. Mol. Sci.*, **2012**, 2, 73.
[38] Qiaohong Su, Adnan Ali Khan, Zhi Su, Chen Tian, Xiaoqin Li, Jiahua Gu, Ting Zhang, Rashid Ahmad, Xintai Su and Zhang Lin, Novel nitrogen-doped KFeS2/C composites for the efficient removal of Cr(VI), *Environ. Sci.: Nano*, **2021**, 8, 1057